# Probing Mobile Charge Carriers in Semiconducting Carbon Nanotube Networks by Charge Modulation Spectroscopy

*Nicolas F. Zorn[1,2], Francesca Scuratti[3], Felix J. Berger[1,2], Andrea Perinot[3], Daniel Heimfarth[1,2], Mario Caironi[3*], and Jana Zaumseil[1,2*]*

[1] Institute for Physical Chemistry, Universität Heidelberg, D-69120 Heidelberg, Germany

[2] Centre for Advanced Materials, Universität Heidelberg, D-69120 Heidelberg, Germany

[3] Center for Nano Science and Technology @PoliMi, Istituto Italiano di Tecnologia, 20133 Milano, Italy

*Email: *zaumseil@uni-heidelberg.de; Mario.Caironi@iit.it*

ABSTRACT

Solution-processed networks of semiconducting, single-walled carbon nanotubes (SWCNTs) have attracted considerable attention as materials for next-generation electronic devices and circuits. However, the impact of the SWCNT network composition on charge transport on a microscopic level remains an open and complex question. Here, we use charge-modulated absorption and photoluminescence spectroscopy to probe exclusively the mobile charge carriers in monochiral (6,5) and mixed SWCNT network field-effect transistors. Ground state bleaching and charge-induced trion absorption features, as well as exciton quenching are observed depending on applied voltage and modulation frequency. Through correlation of the modulated mobile carrier density and the optical response of the nanotubes, we find that charge transport in mixed SWCNT networks depends strongly on the diameter and thus bandgap of the individual species. Mobile charges are preferentially transported by small bandgap SWCNTs especially at low gate voltages, whereas large bandgap species only start to participate at higher carrier concentrations. Our results demonstrate the excellent suitability of modulation spectroscopy to investigate charge transport in nanotube network transistors and highlight the importance of SWCNT network composition for their performance.

KEYWORDS

single-walled carbon nanotubes, networks, charge transport, charge modulation spectroscopy, photoluminescence, trion

Semiconducting single-walled carbon nanotubes (SWCNTs) have emerged as a promising material for future electronic applications as they combine high charge carrier mobilities with mechanical flexibility and solution-processability.[1, 2] Stimulated by the major progress in sorting techniques such as gel chromatography,[3, 4] density gradient ultracentrifugation,[5] aqueous two-phase separation,[6] and polymer-wrapping,[7-9] the reproducible fabrication of high-performance field-effect transistors (FETs) and circuits based on networks of purely semiconducting SWCNTs has become feasible.[10-16] Nonetheless, charge transport in nanotube networks is not yet fully understood especially regarding mixed networks of nanotube species with varying compositions.[17, 18] A detailed understanding of the fundamental transport parameters is necessary to further optimize effective carrier mobilities for competitive network devices at a minimum cost for purification.

Charge transport in semiconductors is commonly studied through temperature-dependent measurements of conductivities and carrier mobilities.[17, 19-21] However, these techniques cannot distinguish between different SWCNT species and thus are not suitable to examine the chirality-dependent contributions to the macroscopic device performance. Given the high sensitivity of SWCNT absorption and emission features to charge carriers, which was shown in several studies,[22-25] electro-optical methods could provide additional and even chirality-specific insights. For example, based on the analysis of the $E_{11}$ absorption band change of (6,5) SWCNTs upon electrochemical doping, Eckstein *et al.* suggested that charges become localized through interaction with counterions.[26] Recently, Ferguson *et al.* investigated redox-doped (6,5) SWCNT films using a combination of quasi-optical microwave and electrical conductivity measurements to quantify the ratio between localized and delocalized carriers.[27] Only few reports, however, have focussed on the chirality and diameter dependence of charge accumulation and transport in networks consisting of different SWCNT species. For nanotube networks composed of five different semiconducting species, static photoluminescence quenching experiments indicated the preferential accumulation of charge carriers on small bandgap SWCNTs.[28] By using gate voltage-dependent electroluminescence (EL), Rother *et al.* investigated preferential current pathways in mixed semiconducting SWCNT networks.[29] The emission from SWCNTs with small bandgaps was found to dominate the EL spectra even though those nanotubes represented a minority species within the network. These experimental results were in good agreement with the carrier density dependent distribution of current among nanotubes obtained by numerical simulations based on a random resistor network.[18] However, the mentioned methods rely on changes in the absorption or emission properties that are static and could be the result of either mobile or trapped charges. Hence, their correlation with actual charge transport in a device is ambiguous. A spectroscopic technique is required that only responds to mobile charge carriers and thus directly provides their optical signatures and distribution in a mixed network of nanotubes; and that can be applied directly to conventional thin film devices.

A highly sensitive method to study charge transport in organic semiconductors is charge modulation spectroscopy (CMS). CMS is a lock-in-based technique that records the differential change in absorption of a semiconducting layer upon modulation of the gate voltage in an operating FET.[30, 31] Since trapped charges cannot follow the bias modulation, the detected spectral features are directly associated with only mobile carriers. The technique was applied successfully to polymer[32-38] and small molecule[39, 40] semiconductors and enabled the identification of different polaron absorption features associated with charge localization and specific molecular orientations within the channel.[41, 42] For ambipolar semiconductors both types of carriers (*i.e.*, electrons and holes) can be studied by changing the polarity of the applied gate voltage as shown for polyselenophene and donor-acceptor copolymer FETs.[36, 43] Through the combination of CMS with a confocal microscopy setup, Chin *et al.* were even able to visualize the ambipolar transport regime and map the distribution of holes and electrons across the FET channel.[36] In a recent study, Pace *et al.* could demonstrate that electrons and holes selectively bleach different absorption transitions in a donor-acceptor copolymer film, thus indicating different interactions and transport properties.[44] Despite these advances in understanding of transport properties of organic semiconductors enabled by CMS, several aspects of the technique have not been explored yet. For example, although the gate voltage modulation is at the heart of CMS, so far only very few reports have put any emphasis on the impact of the applied gate bias or its modulation frequency (typically 37 Hz) on the detected signal and the possible insights gained from that.[30, 31, 43] Most importantly CMS has not yet been applied to SWCNT networks despite their almost ideal optical properties for this technique, such as strong and narrow absorption and emission features and excellent photostability.

Here, we demonstrate charge modulation absorption and photoluminescence spectra of SWCNT FETs and thus probe hole and electron transport in random semiconducting nanotube networks with defined compositions. The dense SWCNT networks and their excellent performance in field-effect transistors resulted in charge-modulated spectra with high signal-to-noise ratios. The achieved sensitivity allowed for frequency-dependent measurements up to modulation frequencies of 100 kHz, only restricted by the detection limit of the lock-in amplifier. Evaluation of the voltage dependence of the modulated $E_{11}$ absorption and emission features provided clear evidence that at low gate voltages mobile charge carriers are preferentially accumulated on and transported through SWCNTs with the smallest bandgaps (largest diameters), even if these only represent a small proportion of the network. We find that the distribution of mobile carriers within the network changes with applied gate voltage and carrier density, thus creating a semiconducting channel that is complex and dynamic in its effective composition.

RESULTS AND DISCUSSION

To study the charge transport in SWCNT networks, we employed selective polymer-wrapping to prepare two different dispersions containing only semiconducting nanotube species. A nearly monochiral dispersion of (6,5) SWCNTs (bandgap 1.27 eV, diameter 0.76 nm) in toluene was obtained from CoMoCAT raw material with poly[(9,9-dioctylfluorenyl-2,7-diyl)-*alt-co*-(6,6'-(2,2'-bipyridine))] (PFO-BPy) as a wrapping polymer *via* shear force mixing as described previously.[45] The absorbance spectrum (**Figure 1a**) shows the characteristic $E_{11}$ transition of (6,5) SWCNTs at 998 nm and the phonon sideband (PSB) at 859 nm. A dispersion of HiPco nanotubes with poly(9,9-dioctylfluorene) (PFO) in toluene yielded a mixture of five chiralities, namely (7,5), (7,6), (8,6), (8,7), and (9,7) SWCNTs with bandgaps from 1.21 to 0.94 eV, as reported by Nish *et al*.[9] Further details on the sample preparation can be found in the **Methods** section. Five distinct peaks corresponding to the $E_{11}$ absorptions are visible in the near-IR (nIR) absorbance spectrum (**Figure 1b**). The abundance of individual nanotube species was determined from fits of the $E_{11}$ absorption peaks and spectrally integrated molar absorptivities[46] (see **Supporting Information, Table S1**). Raman spectroscopy and photoluminescence excitation-emission maps confirmed the absence of significant amounts of other nanotube species (see **Supporting Information, Figures S1 and S2**). **Figure 1c** shows the linear density of states (DOS) of the different species weighted by their respective abundance in the dispersion, indicating the differences between the conduction band energy levels.

Both dispersions were used to fabricate bottom-contact, top-gate FETs (schematic device layout shown in **Figure 1d**). Random SWCNT networks were deposited on photolithographically patterned, interdigitated gold source/drain electrodes with 20 μm channel length. Atomic force microscopy images showed dense networks with well-resolved individual nanotubes, suggesting only negligible residual amounts of unbound polymer (see

**Supporting Information, Figure S3**). A hybrid dielectric, consisting of ~11 nm poly(methyl methacrylate) (PMMA) and a ~61 nm layer of hafnium oxide (HfO$_x$) on top, was chosen to achieve low gate leakage currents and to permit low-voltage operation.[47] To complete the devices, a semi-transparent 20 nm silver gate electrode was thermally evaporated onto the dielectric through a shadow mask.

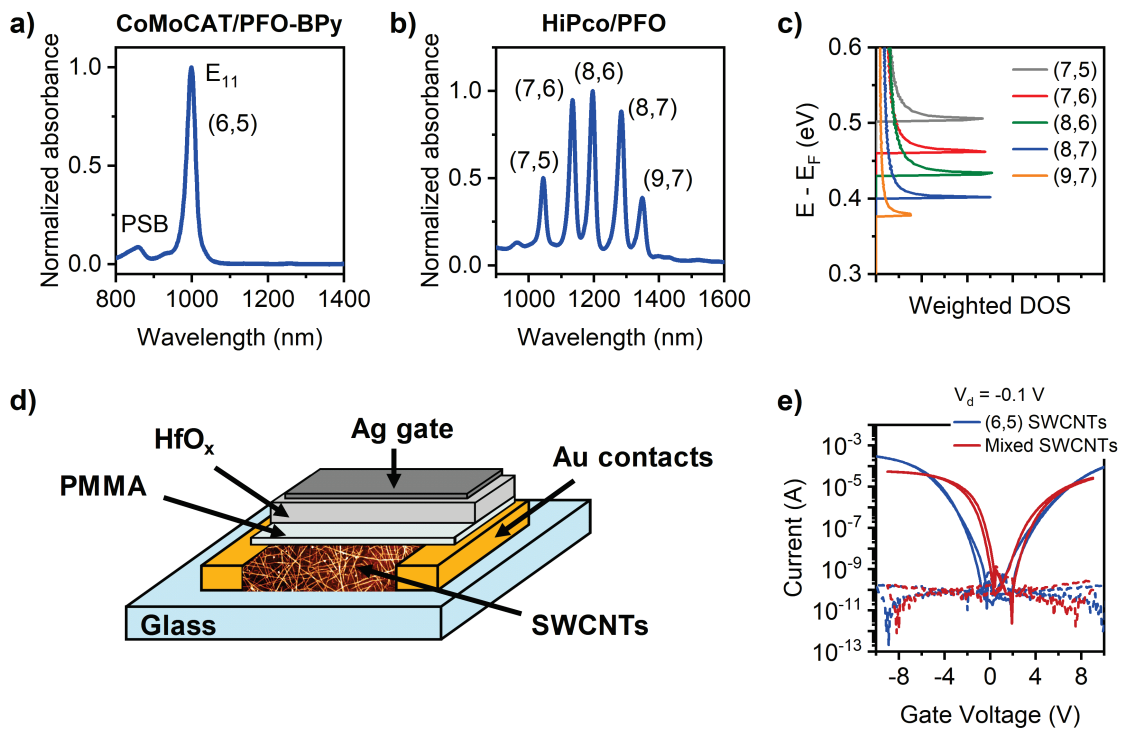

**Figure 1. (a)** Absorbance spectrum of a monochiral (6,5) SWCNT dispersion obtained from CoMoCAT raw material *via* polymer-wrapping with PFO-BPy. **(b)** Absorbance spectrum of a mixed HiPco/PFO dispersion. **(c)** DOS of the five SWCNT chiralities present in the HiPco/PFO dispersion weighted by their abundance. Only the first subbands in the conduction band are shown. **(d)** Schematic structure of SWCNT FETs studied in this work. **(e)** Typical ambipolar transfer characteristics (source-drain current, solid lines; gate leakage, dashed lines) in the linear regime ($V_d$ = -0.1 V) of FETs based on random networks of monochiral (6,5) SWCNTs (blue) and mixed SWCNTs (red), respectively.

All devices exhibited ambipolar charge transport with low leakage currents and high on/off ratios on the order of $10^5$ - $10^6$ as shown in the transfer characteristics at low source-drain bias ($V_d$ = -0.1 V, **Figure 1e**) and the respective output curves (see **Supporting Information**, **Figure S4**). On-currents for the mixed network FETs were lower than for devices based on monochiral (6,5) SWCNT networks. This difference was also reflected in the linear carrier mobilities. For monochiral (6,5) SWCNT networks, average mobilities of 8.2 cm$^2$ V$^{-1}$ s$^{-1}$ for electrons and 9.9 cm$^2$ V$^{-1}$ s$^{-1}$ for holes were obtained. Mixed networks exhibited carrier mobilities of 1.9 cm$^2$ V$^{-1}$ s$^{-1}$ for electrons and 2.3 cm$^2$ V$^{-1}$ s$^{-1}$ for holes. In agreement with recent results, the different mobilities might be attributed to the variation of bandgaps in the mixed network and the associated energy barriers for charge hopping between different SWCNTs.[19,29] Small bandgap nanotubes could potentially act as traps, thus limiting the device performance. The shape of the output curves (**Supporting Information**, **Figure S4**) also indicates a higher injection barrier for electrons than holes and thus contact resistance.[19] However, all devices were suitable for optical characterization in both electron and hole accumulation regimes.

**Monochiral (6,5) SWCNT Networks**

As a model system for the application of charge modulation spectroscopy to investigate nanotube network FETs we chose monochiral (6,5) SWCNTs that are available in large quantities and high quality.[45] All CMS measurements were carried out in transmission as depicted in **Figure S5** of the **Supporting Information**. Understanding the spectral features that correspond to mobile charge carriers in a monochiral SWCNT layer and their gate voltage dependence is the prerequisite for the analysis of mobile carriers in more complex mixed networks. **Figure 2a** shows the CMS spectra of a (6,5) SWCNT network transistor acquired in hole accumulation for different offset (gate) voltages $V_{os}$ with a peak-to-peak modulation $V_{pp}$ of 0.2 V and a modulation frequency of 363 Hz. The complete dataset is shown in **Figure S6a,b** of the **Supporting Information**. Notably, the detected differential change in transmission ($\Delta T/T = 10^{-4} - 10^{-3}$) was three to four orders of magnitude below the static transmission, which underlines the high sensitivity of this method. Two main bleaching signals ($\Delta T/T > 0$) were observed at ~1000 nm and ~870 nm, which match the ground state absorptions of the $E_{11}$ exciton and the PSB, respectively.[48] At low voltages, a charge-induced absorption feature ($\Delta T/T < 0$) emerged at ~1165 nm that can be attributed to absorption into a trion state (charged exciton) in good agreement with previously reported values.[23, 24] The same features were observed for CMS spectra in electron accumulation (**Figure 2b**, full dataset shown in the **Supporting Information**, **Figure S6c,d**), although higher voltages had to be applied to achieve the same bleaching effect. This offset can be attributed to the higher threshold voltage for electron injection probably due to contact resistance, as indicated by the transfer and output characteristics. The large absolute CMS signal and the high signal-to-noise ratio highlight the outstanding suitability of this modulation technique for the investigation of carbon nanotube networks due to their high oscillator strength, photostability and sensitivity toward doping[22] as well as the optimized device structure with a thin but robust high-capacitance gate dielectric that prevents optical interference issues.[47]

The strong and reproducible CMS signal for both hole and electron accumulation in (6,5) SWCNT networks enabled a detailed investigation of the impact of the total carrier density and modulation frequency on the observed spectral features, which has been a challenging task for organic semiconductors. The variation of the offset voltage indeed led to distinct changes in the CMS spectra. All bleaching and charge-induced absorption signals initially increased with hole accumulation until they reached their maximum values at $V_{os} \approx -1.0$ V. This gate voltage corresponds to a carrier density of about $5 \cdot 10^{11}$ cm$^{-2}$. Further increasing the negative offset voltage led to a decrease in intensity of these signals. Eventually, a change from negative to positive ΔT/T at the wavelength of the trion absorption was observed. CMS spectra for electron accumulation revealed the same trends (**Figure 2b**). Notably, the ratio between $E_{11}$ and PSB bleaching signals remained constant throughout all spectra while the $E_{11}$ signal exhibited a blueshift with increasing $V_{os}$, *i.e.* a decrease in exciton binding energy.[49] **Figure 2c** shows the $E_{11}$ peak positions and the extracted CMS signals at the wavelengths of the $E_{11}$ and trion transitions depending on offset voltage. It should be emphasized again that CMS is a differential method and the spectra essentially represent the difference in transmission upon modulation of the mobile carrier density. The observed voltage dependence of the $E_{11}$ signal can be rationalized by taking the device capacitance (increasing with $V_{os}$) and the oscillator strength (decreasing with $V_{os}$) into account. Once the threshold voltage for mobile carrier injection is reached, the steep increase in capacitance with $V_{os}$ leads to a higher modulated charge density and consequently a stronger CMS signal, as shown by the first three data points in **Figure 2c**. A representative capacitance-voltage (C-V) sweep for a (6,5) SWCNT network is shown in the **Supporting Information, Figure S7a**. With increasing doping level, however, the oscillator strength of the $E_{11}$ transition decreases due to static exciton bleaching,[22, 50] whereas the capacitance essentially remains constant once the semiconductor layer is fully charged. Eckstein *et al.* recently quantified the dependence of the $E_{11}$ exciton bleaching on the doping level for electrochemically and chemically doped (6,5) SWCNTs.[50] When plotted linearly, their data show a steep increase in exciton bleach for low doping levels, which flattens for high charge densities. The derivative of the exciton bleach - which is the quantity to compare due to the differential nature of modulation spectroscopy - steadily decreases with increasing carrier densities. This decrease is reflected in the attenuation of the $E_{11}$ bleaching signal intensity in the CMS spectra (electrostatic doping) at higher offset voltages. The data corroborate that at high $V_{os}$, the $E_{11}$ absorption is bleached to such an extent that the differential change in transmission due to gate modulation becomes less and less significant.

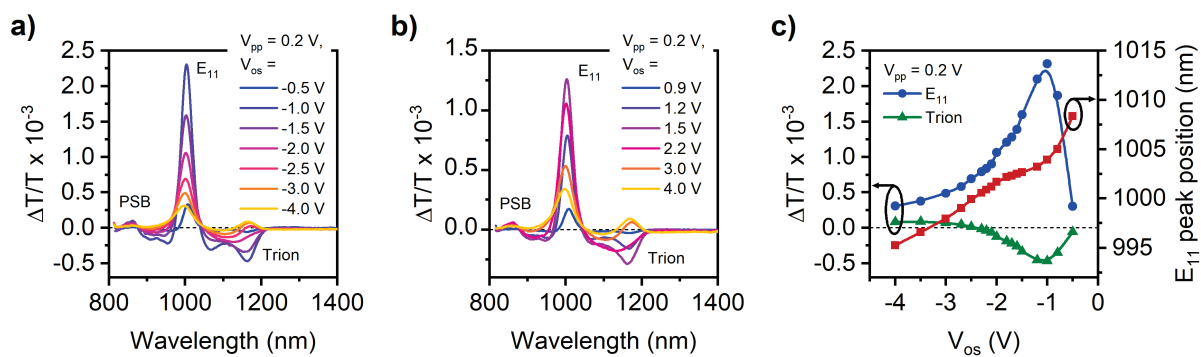

**Figure 2.** $V_{os}$-dependent CMS spectra ($V_{pp}$ = 0.2 V) of (6,5) SWCNTs **(a)** in hole accumulation and **(b)** in electron accumulation. **(c)** Extracted CMS signals of (6,5) SWCNTs (hole accumulation) at the wavelengths of $E_{11}$ bleaching and trion absorption (left axis) and $E_{11}$ peak positions (right axis). Lines are guides to the eye.

The charge-induced absorption feature corresponding to the trion decreases in intensity and even changes sign from negative to positive ΔT/T with increasing offset voltage. Previous studies on static voltage-dependent absorbance of SWCNTs reported the so-called heavily doped regime that is characterized by a broad, featureless absorption band (denoted as "*H*-band") across the visible and nIR spectral regions between approximately 630 nm and 1170 nm.[22, 23] The trion optical density reaches a maximum as a function of doping and subsequently decreases until only a broad *H*-band remains.[22, 23, 26, 27] This decrease is reflected in a positive ΔT/T signal (*i.e.*, trion bleaching) in the CMS spectra and thus explains the observed evolution of the trion signature.

We compared our CMS spectra to the absorbance spectra of a (6,5) SWCNT film in an electrochromic device at different voltages, as reported by Berger *et al.*[23] (see **Supporting Information, Figure S8**). The subtracted static spectra clearly have a very similar shape to the CMS spectra and reveal identical trends for an increasing total charge density, *i.e.* decrease of the $E_{11}$ bleaching signal and change from a charge-induced absorption to a bleaching signal for the trion absorption. Furthermore, the comparison to the spectroelectrochemical data corroborates that the small, negative dips next to the $E_{11}$ bleaching signal, which are also visible in CMS, are due to the spectral evolution upon doping of the SWCNT film and not due to electroabsorption. Note that in order to ensure the correct assignment of the observed transitions and to prevent misinterpretation of any features, additional cross-checks (second harmonic detection and measurements under different angles of the incident light beam) were performed. A detailed discussion and second harmonic spectra can be found in the **Supporting Information, Figure S9**.

Furthermore, we performed CMS measurements in the visible spectral range and observed the charge-induced bleaching of the $E_{22}$ transition at ~580 nm. The simultaneous modulation of the two absorption features corroborates a common ground state and the excitonic nature of the excited state of SWCNTs. However, the $E_{22}$ / $E_{11}$ ratio of the CMS signals is considerably lower than that of the static absorbance spectrum. This discrepancy in the decrease of oscillator strengths with voltage has been previously reported for electrochemically doped SWCNT films.[22, 23] A spectrum comprising both the visible and the nIR regions is shown in the **Supporting Information**, **Figure S10**. The absence of any spectral features that could be ascribed to polymer polaron bands corroborates that the wrapping polymer does not participate in charge transport in networks of polymer-sorted SWCNTs.

So far, all CMS spectra were obtained by modulating the gate voltage with a frequency of 363 Hz. In order to study the temporal response of the mobile charge carriers, frequency-dependent measurements were performed on (6,5) SWCNT network transistors. **Figure 3a** shows a dependence on the modulation frequency in the CMS spectra obtained at moderate hole accumulation ($V_{os}$ = -1.5 V). The CMS signal decreased by approximately a factor of two upon increase of the frequency by one order of magnitude (3035 Hz), and by a factor of ten when modulating at 36295 Hz. The normalized spectra, however, are essentially identical (see **Supporting Information, Figure S11a**), corroborating that the signals have a common physical origin, *i.e.* they are solely charge-induced. CMS signals with different physical origins are expected to respond to the voltage modulation on different timescales.[30] This notion was used in previous CMS experiments on oligothiophenes and poly(3-hexylthiophene) to distinguish between polarons and bipolarons.[51, 52] For high carrier concentrations ($V_{os}$ = -3.5 V, **Figure 3b**), the CMS spectra of the (6,5) SWCNT FETs showed almost no dependence on the modulation frequency up to the detection limit of the lock-in amplifier (100 kHz). The normalized spectra in the **Supporting Information, Figure S11b**, are identical.

The variation in frequency dependence with $V_{os}$ can be rationalized with the linear carrier mobilities. For low gate (offset) voltages ($|V_g| < 2$ V), effective carrier mobilities are still low (below $10^{-2}$ cm$^2$ V$^{-1}$ s$^{-1}$) as the device is not yet in full accumulation (for a graph of the voltage-dependent mobility, see **Figure S12a** of the **Supporting Information**). Hence, charges cannot respond fully at higher modulation frequencies and the CMS signal decreases. For higher $V_g$, the mobilities are sufficiently high (> 1 cm$^2$ V$^{-1}$ s$^{-1}$) for carriers to follow modulation frequencies of almost 100 kHz, resulting in nearly frequency-independent CMS features. This difference in cut-off with gate voltage is also shown from frequency-dependent capacitance measurements (see **Supporting Information, Figure S13**). Further evidence is provided by the frequency-dependent device admittance as measured by modulating the source terminal of the FET (**Figure 3c**). Such frequency-dependent measurements can be employed to evaluate the physical limit to the device operation frequency ($f_\tau$) as imposed by the carrier transit time.[53] For $f < f_\tau$, the carrier velocity is sufficient to follow the modulation, yielding a constant admittance value. At higher frequencies, however, the admittance will drop due to the limited carrier velocity. The data in **Figure 3c** and the corresponding phase plot in **Figure S14** (**Supporting Information**), shows that $f_\tau$ is above 100 kHz for full accumulation ($V_g$ = -4 V), while it is in the range of few kHz at lower gate voltages ($V_g$ = -1 V). These data are in good agreement with the frequency-dependent CMS spectra, whose maximum frequency was limited by the lock-in amplifier when in full accumulation (see **Figure 3b**).

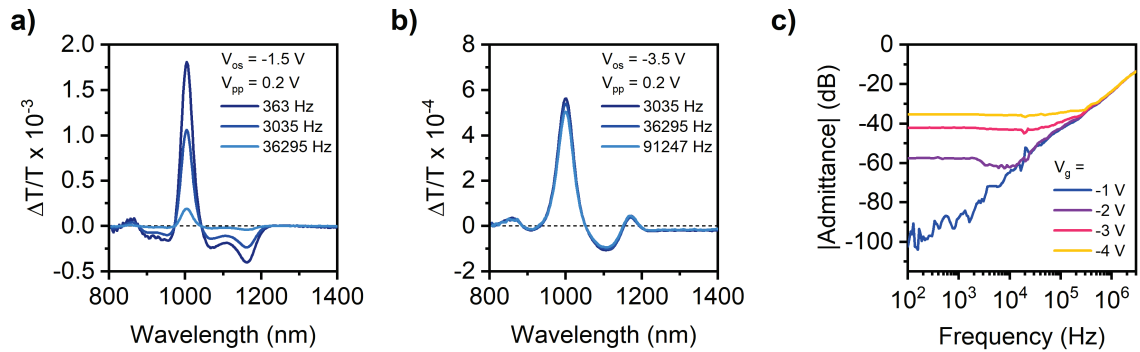

**Figure 3.** Frequency-dependent CMS spectra of (6,5) SWCNTs **(a)** at moderate hole doping level ($V_{os}$ = -1.5 V, $V_{pp}$ = 0.2 V) and **(b)** in strong hole accumulation ($V_{os}$ = -3.5 V, $V_{pp}$ = 0.2 V). **(c)** Frequency-dependent magnitude of the device admittance of a (6,5) SWCNT FET for different gate voltages $V_g$. The contribution scaling with modulation frequency is due to the source-to-drain capacitance.

As a complementary method to investigate charge transport in SWCNT networks, we turn to the nanotube photoluminescence (PL), which is even more sensitive to injected charges due to nonradiative Auger quenching of excitons.[54, 55] Static voltage-dependent PL spectra were acquired with the (6,5) SWCNTs being excited resonantly at the $E_{22}$ transition at 577 nm. **Figure 4a** shows a steady decrease in intensity of the $E_{11}$ emission feature and the redshifted PSB with increasing gate voltage and thus carrier density. Trion emission becomes visible at 1180 nm and eventually exceeds the exciton PL intensity at high carrier concentrations. The trion peak is shifted by 176 meV with respect to the exciton in good agreement with the approximation by Park *et al.*[56] For electron accumulation (data shown in **Figure S15** of the **Supporting Information**), an energy difference of 181 meV between exciton and trion emission was observed, which is also consistent with previous reports on the energy shifts for positive and negative trions, respectively.[24] The $E_{11}$ emission peak blueshifts with increasing $V_g$ for both hole and electron doping similar to the CMS $E_{11}$ bleaching peak although to a much lesser extend (approximately 3 nm *versus* 13 nm).

Emission quenching in these static PL measurements occurs through interaction of excitons with both mobile and trapped charges. Thus, it does not provide unambiguous information on the distribution of mobile carriers that contribute to the transistor current. To address this issue, we added the modulation technique to our PL setup in order to perform charge modulation PL spectroscopy (CMPL). This technique combines the high sensitivity of the SWCNT emission to injected charges and the specificity of modulation spectroscopy to only mobile carriers. A schematic of the measurement setup can be found in the **Supporting Information, Figure S16**.

In CMPL, the gate voltage is again modulated with a sinusoidal peak-to-peak voltage $V_{pp}$ around an offset voltage $V_{os}$. During this modulation, the sample is excited with monochromatic laser light and the differential change in PL at the frequency of modulation is recorded with a lock-in amplifier. Only the low-voltage regime is accessible in CMPL measurements due to the highly efficient, charge-mediated PL quenching and thus loss of emission intensity. In agreement with the data obtained from CMS and frequency-dependent admittance measurements (**Figure 3**), a frequency dependence of the (6,5) SWCNT CMPL signal intensity was observed, but no change in the normalized signal (**Supporting Information, Figure S17**). Hence, a modulation frequency of 73 Hz was chosen for all experiments.

CMPL spectra ($V_{pp}$ = 0.3 V) at different offset voltages for an FET with monochiral (6,5) SWCNTs are shown in **Figure 4b**. The differential PL signal (ΔPL; positive values indicate quenching, whereas negative signals indicate charge-induced emission) corresponding to $E_{11}$ exciton quenching shows a similar behaviour to that in the CMS experiments. Initially, the signal increases with $V_{os}$ due to the increase of modulated charge carrier concentration. The maximum ΔPL intensity was observed at $V_{os} \approx$ -1.0 V, which is in good agreement with the value extracted from CMS. For higher voltages, the CMPL signal decreases again. In contrast to the experiments in transmission no distinct trion feature was observed. This is consistent with the static absorbance and PL spectra (**Figures S8a** and **4a**), which show that charge-induced trion absorption is already very pronounced at low voltages, whereas trion emission only becomes significant at high carrier densities. Since modulation spectroscopy records the change in absorption/emission upon carrier modulation, the trion signal is basically absent in the CMPL spectra. In a different approach, we simultaneously varied $V_{os}$ and $V_{pp}$ in order to modulate between zero bias (off-state) and on-state ($V_{pp}$ = 2 $|V_{os}|$) in analogy to experiments by Koopman *et al.* on PL electro-modulation microscopy of organic FETs.[57] This measurement scheme consists of a stepwise increase of the modulated charge density. Thus, with increasing voltage, the $E_{11}$ ΔPL signal should approach the total PL intensity of the unbiased film, as indicated in **Figure 4c**. The PL of the unbiased film (shown for comparison) was obtained by modulating the excitation laser with a mechanical chopper that simultaneously served as reference for the lock-in amplifier. In **Figures 4a** and **4c**, the wavelength of trion emission is indicated with an arrow. As shown for the static experiments (**Figure 4a**), the trion PL intensity is comparable to the intensity of the redshifted PSB in the voltage range probed with CMPL, and is only slightly higher for high voltages (*e.g.*, -5 V). Due to the differential nature of CMPL, this is reflected in a dip in the ΔPL signal at the trion wavelength.

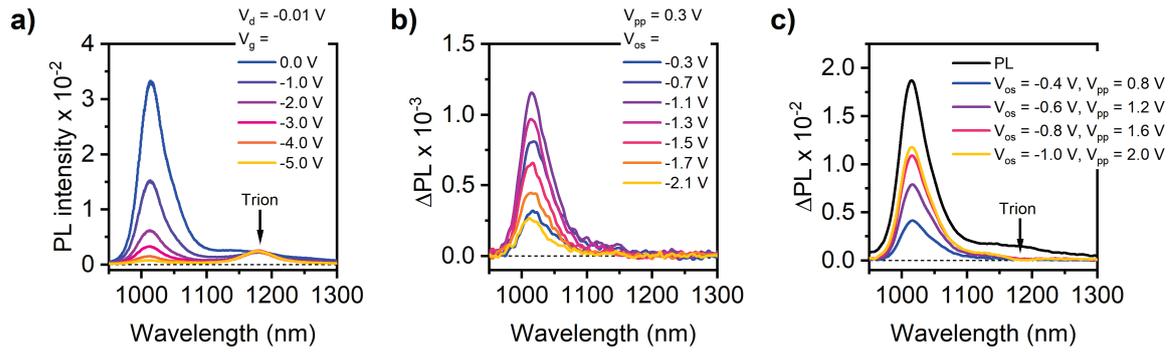

**Figure 4. (a)** Static gate voltage-dependent PL spectra and **(b)** $V_{os}$-dependent CMPL spectra ($V_{pp}$ = 0.3 V) of (6,5) SWCNTs in hole accumulation. **(c)** CMPL spectra of (6,5) SWCNTs with modulation between zero bias and on-state of the FET.

Overall, both techniques – CMS and CMPL – are suitable to quantitatively follow spectral features associated with mobile charge carriers in SWCNT network FETs. The $E_{11}$ bleaching and quenching features depend strongly on the concentration of mobile charge carriers. The trion absorption, which is very prominent in the CMS measurements, changes substantially with offset voltage and may complicate the analysis of CMS signals from mixed nanotube networks. For those samples the very weak charge-induced trion emission in the CMPL spectra compared to the $E_{11}$ emission quenching should be advantageous.

**Mixed SWCNT Networks**

After clarifying the different spectral contributions and their voltage dependence for (6,5) SWCNT FETs, we now turn to the mixed SWCNT networks based on the HiPco/PFO dispersion, consisting of five semiconducting nanotube species as shown in **Figure 1b**. Through comparison with the static absorbance spectrum, the five dominant bleaching features in the $V_{os}$-dependent CMS spectra in **Figure 5a** ($V_{pp}$ = 0.2 V, modulation frequency 363 Hz) were assigned to the $E_{11}$ transitions of the different nanotube species. All $E_{11}$ bleaching peaks show a blueshift with increasing $V_{os}$. For the (8,7) and (9,7) SWCNTs the trion features are visible as a broad charge-induced absorption band at wavelengths above 1400 nm. Trion absorption peaks corresponding to the other nanotube species are difficult to identify due to the spectral overlap. Additional CMS spectra with 2ω detection (**Supporting Information, Figure S18**) did not show any significant contributions from electroabsorption similar to the (6,5) SWCNT devices.

In agreement with the voltage-dependent spectral evolution of the $E_{11}$ feature of the monochiral samples (see above), the exciton bleaching signals for the mixed networks initially increased to a maximum intensity and then decreased. Again, this change can be attributed to the initial increase of the device capacitance (for a C-V double sweep of a mixed network FET, see the **Supporting Information, Figure S7b**) followed by a reduced total oscillator strength of the $E_{11}$ transition. In comparison to (6,5) SWCNTs, however, the gate (offset) voltages for maximum bleaching as well as the highest voltage to acquire a spectrum were significantly lower. This difference results from the smaller bandgaps of the five nanotube species in the mixed network compared to the (6,5) SWCNTs. A closer inspection of the spectra also revealed a pronounced difference between the CMS peak intensities at the lowest offset voltage ($V_{os}$ = -0.35 V) and the static absorbance spectrum (**Figures 1b** and **5a**). The (8,7) SWCNT exciton bleaching is dominant, whereas only small CMS signals can be attributed to (7,5) and (7,6) nanotubes. Furthermore, the signal intensities change relatively to each other with $V_{os}$, which is more clearly shown in the CMS spectra normalized to the (8,6) $E_{11}$ bleaching feature (**Figure 5b**). With increasing offset voltage, bleaching signals of the nanotubes with larger bandgaps increase with respect to the (8,6) peak, whereas those of smaller bandgap SWCNTs decrease.

It should be emphasized again that only mobile (not trapped) charges are detected in CMS. Hence, these measurements allow us to draw direct conclusions on the distribution of mobile carriers within the complex electronic structure of mixed nanotube networks and thus charge transport within them. At low voltages, small bandgap SWCNTs carry the largest portion of mobile charges leading to high relative bleaching signals. With increasing $V_{os}$, the relative contribution of the large bandgap nanotubes to the overall charge transport rises and thus a relative increase in the respective signal intensities is observed. Meanwhile the CMS bleaching signals corresponding to the small bandgap SWCNTs already decrease again. These observations are consistent with previous results from voltage-dependent electroluminescence spectroscopy and theoretical simulations of HiPco/PFO networks, which suggested that the distribution of charges within the network changes with gate voltage.[18, 29]

**Figure 5c** shows the extracted CMS signal intensities at the wavelengths of the different $E_{11}$ transitions normalized to their respective maxima. The bleaching intensities for the individual species show the same general evolution as the (6,5) SWCNTs (**Figure 2c**) but reach their maxima at different $V_{os}$. Note that the evaluation of the (9,7) SWCNT $E_{11}$ bleaching was not possible due to the spectral overlap with the charge-induced trion absorption of the (8,6) nanotubes. The offset voltage values for the bleaching maxima follow the energetic levels of the valence band and thus bandgaps of the nanotubes (see **Figure 1c**). The larger the bandgap, the higher the offset voltage at which the highest CMS intensity was recorded. Furthermore, the drop of intensity with $V_{os}$ is steeper for the small bandgap nanotubes compared to the large bandgap SWCNTs. We can conclude that the distribution of mobile carriers within mixed SWCNT networks depends strongly on the overall carrier density and the distribution of nanotube species with different bandgaps. A disproportionate amount of current, *i.e.* more than what is expected from the relative amounts of SWCNT species within the network, is carried by the small bandgap nanotubes especially at lower gate voltages. At high gate voltages, when the device is in full accumulation, the entire network appears to contribute to the charge transport.

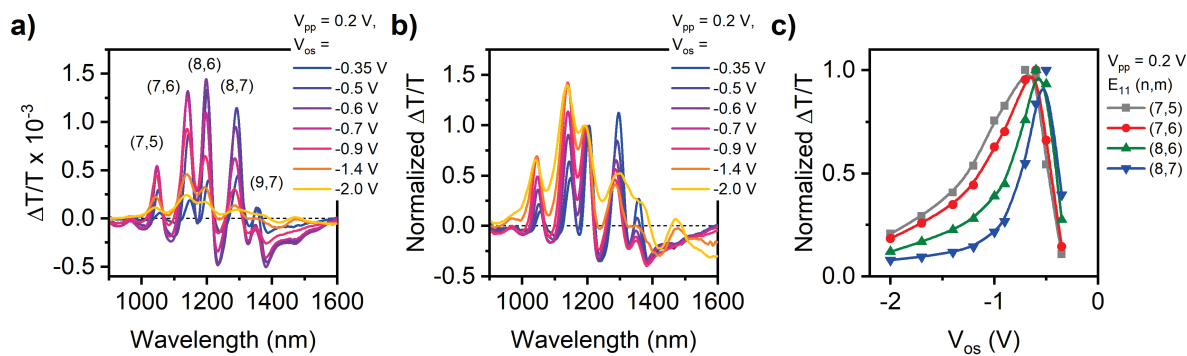

**Figure 5. (a)** $V_{os}$-dependent CMS spectra of mixed SWCNTs in hole accumulation ($V_{pp}$ = 0.2 V). **(b)** Spectra normalized to the (8,6) SWCNT $E_{11}$ bleaching signal. **(c)** Extracted CMS signals of mixed SWCNTs in the hole accumulation regime ($V_{pp}$ = 0.2 V) normalized to the maximum bleaching intensity. Lines are guides to the eye.

Similar to the monochiral (6,5) SWCNTs, a strong frequency dependence of the CMS signal was observed for the mixed network FETs at low voltages (**Supporting Information, Figure S19a,b**). At higher frequencies the CMS signal was again lower but the normalized spectra and peak ratios were essentially identical, confirming the common physical origin of the observed features. The frequency- and gate voltage-dependent admittance of mixed SWCNT FETs (**Supporting Information, Figure S19c**) was also very similar to that of (6,5) SWCNT-based devices. A graph of the voltage-dependent mobility is shown in **Figure S12b** of the **Supporting Information**.

To further corroborate the notion of chirality-dependent charge accumulation and transport, we investigated the PL properties of the mixed networks. Successive, static voltage-dependent PL quenching with increasing gate voltage is shown in **Figure 6a**. Evidently, emission from the small bandgap SWCNTs is strongly quenched, whereas the PL from (7,5) SWCNTs is only weakly affected. This observation of chirality-dependent accumulation of carriers within the nanotube network is in very good agreement with previous experimental results as well as the expected static carrier density distribution based on a one-dimensional semiconductor model.[28, 29] Notably, trion emission becomes significant at higher voltages (peaks at 1360 nm and 1480 nm).

To exclusively probe mobile charge carriers, we performed CMPL experiments on mixed SWCNT FETs. Despite the decrease in signal intensity with increasing modulation frequency, frequency-dependent measurements confirmed that the peak intensities remained constant in relation to each other (see **Supporting Information, Figure S20**). By modulating between the off- and on-state of the device, the gradual increase in total PL quenching resulting from the stepwise increase in modulated charge density was recorded. As evident from the spectra in **Figure 6b**, PL quenching is strongest for the small bandgap (8,7) and (9,7) SWCNTs. For $V_{os}$ = -0.6 V (*i.e.*, modulation between 0 V and -1.2 V), emission from these two species is quenched predominantly, whereas the total PL quenching of (7,5) SWCNTs determined from peak fits is only 24%.

To further quantify the chirality dependence of charge-induced quenching, the fraction of PL quenched by mobile carriers (ΔPL/PL) was calculated for each individual chirality. Assuming that the efficiency of exciton quenching is independent of the SWCNT species, the share of ΔPL/PL (**Figure 6c**) should reflect the fraction of mobile carriers passing through SWCNTs of a given chirality, or in other words, the current share for this chirality. The graphs in **Figure 6c** show that the portion of mobile charge concentration is not proportional to the abundance of the SWCNT species in the network, but is highest for the small bandgap SWCNT species. The data further corroborates that the distribution of mobile carriers changes with applied gate voltage. The contribution of large bandgap SWCNTs to the current increases with $V_{os}$.[18, 29] Schießl *et al.* simulated carrier density-dependent charge transport in mixed HiPco/PFO networks by solving Kirchhoff's equations for a random resistor network.[18] Charge carrier concentrations in these simulations were on the order of $10^{10} - 10^{12}$ cm$^{-2}$, which is comparable to those in the CMPL experiments. The CMPL data shown in **Figure 6c** nicely reproduces the gate voltage-dependent current distribution of the simulated networks with very similar species fractions. Especially the current shares for (7,5) and (7,6) SWCNTs are in excellent agreement (5-10% and 10-20%, respectively). The model predicted a current share of 10-15% for the (9,7) SWCNTs with an abundance of 6%. In CMPL, the ΔPL/PL share was 25-35% (calculated abundance of 10.5%).

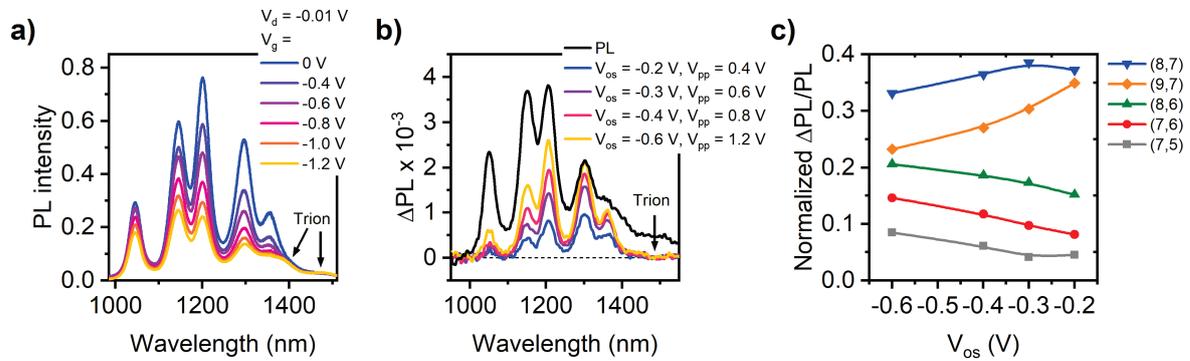

**Figure 6. (a)** Static voltage-dependent PL spectra of a mixed SWCNT network FET and **(b)** CMPL spectra with modulation between zero bias and on-state of the same device. **(c)** Normalized fraction of quenched PL (ΔPL/PL) for each chirality in the mixed SWCNT networks. Values were determined from peak fits and represent the share of mobile carriers on these nanotubes and thus, the current share at varying total charge densities (offset voltages $V_{os}$). Lines are guides to the eye.

To further assess the validity of the CMPL method, we investigated FETs based on mixed HiPco/PFO SWCNT networks with another network composition. Differences in nanotube species abundance (see the **Supporting Information, Table S2**) are due to variations in the polymer-wrapping process. The absorbance spectrum of the dispersion and representative transfer and output characteristics of these devices are shown in the **Supporting Information,**

**Figure S21**. Additional CMPL measurements of FETs in hole and electron accumulation confirmed the nanotube species-dependent PL quenching through modulation of charge density (**Supporting Information, Figure S22**). Despite the different abundances, the overall trends such as the dominant ΔPL/PL contribution by small bandgap SWCNTs and the change of mobile carrier distribution with increasing gate voltage are reproduced. Thus, by assuming that the relative ΔPL/PL intensity reflects the current share of a particular nanotube species, we can infer that charge transport is dominated by the small bandgap SWCNTs within a mixed network at low gate voltages before charges can access and travel through the entire network at high gate voltages.

CONCLUSION

We have studied mobile charge carriers in FETs based on SWCNT networks by means of charge-modulated absorption (CMS) and photoluminescence (CMPL) spectroscopy. The unusually strong signal intensities and high signal-to-noise ratios for these devices enabled a thorough analysis of the modulated spectra depending on total charge density and modulation frequency. In agreement with previous experiments on chemically and electrochemically doped SWCNTs, we observed charge-induced exciton bleaching/quenching and trion absorption/emission features that can now unambiguously be associated with mobile carriers rather than localized or trapped charges. By establishing a clear correlation between the modulated concentration of mobile carriers and the detected signals, we could further corroborate the idea that charge transport in mixed semiconducting SWCNT networks strongly depends on network composition and gate voltage. It occurs preferentially through small bandgap (large diameter) SWCNTs at lower voltages, irrespective of their nominal abundance within the network. The distribution of mobile carriers and thus current changes as a function of applied gate voltage, with large bandgap SWCNTs starting to participate only at higher voltages. This increase of active transport paths with carrier density in a mixed network suggests that typical assumptions and equations for the current-voltage characteristics of FETs with homogeneous semiconducting layers may not be applicable for SWCNT networks at low gate voltages. However, in full accumulation and at high gate voltages the notion that the entire network participates in the transport appears to be valid. Nevertheless, monochiral networks of small bandgap nanotubes do not suffer from these issues and hence are most promising for SWCNT-based FETs with maximum performance. Our results demonstrate the merits of gate voltage- and frequency-dependent modulation spectroscopy of FETs in general, as well as their excellent suitability for the *in-situ* investigation of mobile carriers in SWCNT network devices beyond model systems. Further CMS and CMPL studies could be used to examine the impact of intentional functionalization of nanotubes, for example, with luminescent sp$^3$ defects,[58] on charge transport in networks.

METHODS

**Preparation of (6,5) SWCNT Dispersions.** Nearly monochiral (6,5) SWCNT dispersions were obtained *via* polymer-wrapping using shear force mixing as previously described in detail by Graf *et al.*[45] Briefly, 50 mg of the CoMoCAT SWCNT raw material (CHASM Advanced Materials Inc., Charge No. SG65i-L58) were added to a solution of 65 mg poly[(9,9-dioctylfluorenyl-2,7-diyl)-*alt-co*-(6,6'-(2,2'-bipyridine))] (PFO-BPy, American Dye Source, $M_W$ = 34 kg mol$^{-1}$) in 140 mL of toluene and shear mixed for 72 hours at 20 °C (Silverson L2/Air mixer, 10230 rpm). The dispersion was centrifuged twice for 45 min at 60000$g$ (Beckman Coulter Avanti J26XP centrifuge) and the supernatant was collected. The dispersion was filtered through a polytetrafluoroethylene membrane (Millipore JVWP, pore size 0.1 µm) and the filter cake was washed with hot toluene (80 °C, three times, 10 min each) to remove excess unwrapped polymer. Ultrasonication in 1 mL of toluene for 30 min gave a strongly coloured purple dispersion, which was adjusted to an optical density of 10 cm$^{-1}$ at the wavelength of E$_{11}$ absorption and immediately used for spin-coating of SWCNT networks.

**Preparation of Mixed SWCNT Dispersions.** 15 mg of HiPco SWCNT raw material (Unidym Inc., Batch No. 2172) were added to a solution of poly(9,9-dioctylfluorene) (PFO, Aldrich, M$_W$ > 20 kg mol$^{-1}$) in toluene (10 mL, 2 mg mL$^{-1}$). After ultrasonication for 45 min (Branson 2510 sonication bath), the suspension was centrifuged for 45 min at 60000*g* (Beckman Coulter Avanti J26XP centrifuge) and the supernatant was collected. To increase the yield of dispersed SWCNTs, the pellet was recycled twice and the supernatants were combined. To reduce the content of excess unwrapped polymer, PFO-wrapped SWCNTs were sedimented by ultracentrifugation for 20 h at 284600*g* (Beckman Coulter Optima XPN-80 centrifuge) and the supernatant was removed. The pellets were washed by addition of tetrahydrofuran (THF) and subsequent centrifugation for 30 min at 30000*g*. Bath sonication of the pellet in 1 mL of toluene for 30 min yielded a light green dispersion that was immediately used in the FET fabrication process.

**FET Fabrication.** Interdigitated bottom contact electrodes were patterned on glass substrates (Schott AG, AF32eco, 300 µm) using double-layer resist photolithography (MicroChem LOR5B/Microposit S1813 resist; Suess MicroTec MA6 mask aligner, UV dose 182 mJ cm$^{-2}$) and electron-beam evaporation (Winter Vakuum Technik HVB 130) of chromium (2 nm) and gold (30 nm), followed by lift-off in *N*-methyl pyrrolidone (NMP). Substrates were cleaned by ultrasonication in acetone and 2-propanol for 10 min, respectively, and UV ozone treatment (Ossila E511, 10 min) before deposition of the semiconductor layer. The SWCNTs were deposited either *via* spin-coating or aerosol-jet printing onto the patterned electrodes as detailed below. The samples were annealed at 300 °C for 1 h in dry nitrogen atmosphere, followed by spin-coating (4000 rpm, 60 s) of 80 µL of a solution of poly (methyl methacrylate) (PMMA, Polymer Source, $M_W$ = 315 kg mol$^{-1}$, syndiotactic) in *n*-butyl acetate (6 g L$^{-1}$). On top of the ~11 nm PMMA layer, a ~61 nm HfO$_x$ layer was deposited *via* atomic layer deposition (Ultratech Inc., Savannah S100) at a temperature of 100 °C using a tetrakis(dimethylamino)hafnium precursor (Strem Chemicals Inc.) and water as oxidising agent. Thermal evaporation (MB-ProVap 3G, M. Braun Inertgas-Systeme) of silver top gate electrodes (20 nm) using shadow masks completed the devices.

**SWCNT Network Deposition.** SWCNT networks were deposited either *via* spin-coating or aerosol-jet printing. Spin-coating of SWCNT inks (3× 80 µL, 2000 rpm, 30 s) was performed with annealing steps (100 °C, 2 min) in between. After rinsing with THF and 2-propanol, spin-coated nanotube networks were patterned in order to remove all SWCNTs outside the channel area. This was achieved using photolithography as described above, oxygen plasma treatment (Nordson MARCH AP-600/30, 100 W, 2 min), and lift-off in NMP. Aerosol-jet printing of SWCNTs with an AJ200 printer (Optomec Inc.) followed the procedure described in detail by Rother *et al.*[59] Terpineol (2-5 vol.-%) was added to the SWCNT dispersion to increase viscosity for aerosol formation. A 200 µm inner diameter nozzle was used at a sheath gas and carrier gas (nitrogen) flow of 30 sccm and 25 sccm, respectively. During the process, the movable stage was heated to 100 °C. The substrates were subsequently rinsed with THF and 2-propanol.

**Characterization.** Baseline-corrected absorbance spectra of SWCNT/polymer dispersions were recorded with a Cary 6000i UV-vis-nIR spectrometer (Varian Inc.). Raman spectra on drop-cast SWCNT/polymer films were obtained with a Renishaw inVia confocal Raman microscope in backscattering configuration using an Olympus ×50 long working distance objective (N.A. 0.5) and three different lasers (532 nm, 633 nm, 785 nm). A Bruker Dimension Icon atomic force microscope (AFM) in the ScanAsyst mode was used for the acquisition of AFM images under ambient conditions. Current-voltage measurements (transfer and output characteristics) were performed in dry nitrogen atmosphere with an Agilent 4156C semiconductor parameter analyser. The effective device capacitances were obtained with an impedance analyser (Solatron Analytical ModuLab XM MTS System) under inert atmosphere and with the transistors operated as plate capacitors (source and drain electrodes shorted and grounded). The frequency was 100 Hz and the maximum capacitance was extracted in the on-state of the device. Frequency-dependent admittance measurements were performed in a custom setup under inert atmosphere. An Agilent E5061B Vector Network Analyzer (VNA) provided a sinusoidal signal superimposed on a DC bias to the transistor's source terminal, while the gate electrode was biased independently using an Agilent B2912A source meter. During the measurement, the gate-source DC bias was varied between -1 V and -4 V while the drain-source DC bias was set at 100 mV. The current flowing into the FET drain terminal was collected using a transimpedance amplifier (Femto DHPCA-100), whose output was fed back into the VNA, yielding an admittance function. The amplifier gain was set to $2\times10^3$, the sinusoidal signal was set to a fixed power of 0 dBm, and the maximum signal frequency was set to 2 MHz.

**Charge Modulation Spectroscopy.** For CMS measurements in the transmission mode, the FETs were mounted in a custom-made vacuum chamber (pressure < $10^{-6}$ mbar) with electrical feedthroughs. A schematic of the setup is shown in the **Supporting Information, Figure S5**. Source and drain electrodes were grounded and the gate voltage was modulated with a peak-to-peak voltage $V_{pp}$ around an offset voltage $V_{os}$ using a Keithley 3390 waveform generator. A modulation frequency of 363 Hz was used unless otherwise specified. As the light source, the spectral output of a tungsten-halogen lamp (SP ASBN-W) was wavelength-filtered (step width 3 nm) with a monochromator (SP DK240 1/4m) using gratings with 660/mm ruling, 1200 nm blaze (nIR) and 1200/mm ruling, 550 nm blaze (visible), respectively. The light was focussed on the device (illumination spot size ~1×1 mm²) and the transmitted light was focussed on a photodiode detector (InGaAs, Thorlabs FGA21 (nIR) or Si, Thorlabs FDS100 (visible)). A transimpedance amplifier (Femto DHPCA-100) provided the signal for the lock-in amplifier (Stanford Research Systems SR830), which recorded the differential transmission signal ΔT. The reference transmission spectrum T was acquired using the same setup with no applied gate bias and by modulating the light with an optical chopper (Stanford Research Systems SR540). The ΔT/T CMS spectrum was obtained by dividing the first harmonic (1ω) component of ΔT by the transmission T to correct for the substrate absorption, the absorption of optics in the detection path, and the wavelength-dependent output of the light source.

**Photoluminescence Spectroscopy.** For the acquisition of static gate voltage-dependent PL spectra, the three terminals of the device were electrically connected and a bias was applied to the gate electrode of the device (fixed source-drain bias $V_d$ = -0.01 V) using a Keithley 2612A source meter. Nanotube layers were excited with a supercontinuum laser (Fianium Ltd. WhiteLase SC400, 20 MHz repetition rate, 6 ps pulse width) that was wavelength-filtered at 577 nm ((6,5) SWCNTs) and 600 nm (mixed SWCNTs), respectively, and focussed by a nIR-optimized ×50 objective (Olympus LCPLN50XIR, N.A. 0.65). Emitted photons were collected with the same objective and passed through a dichroic mirror (cut-off wavelength 875 nm) and a longpass filter (LP830) to block scattered light of the laser beam. Spectra were recorded with an Acton SpectraPro SP2358 spectrometer (grating blaze 1200 nm, 150 lines per mm) and a liquid nitrogen-cooled InGaAs line camera (Princeton Instruments OMA V:1024). Each spectrum was corrected to account for the absorption of the optics in the detection path and wavelength-dependent detection efficiency of the camera.

**Charge Modulation Photoluminescence Spectroscopy.** CMPL spectra were acquired with a modified PL spectroscopy setup. For a schematic setup, see the **Supporting Information, Figure S16**. The source and drain electrode were grounded and the gate bias was modulated with a peak-to-peak voltage $V_{pp}$ around an offset voltage $V_{os}$ using a Keysight 33600A waveform generator. Unless otherwise specified, a modulation frequency of 73 Hz was used. The PL was spectrally resolved (step width 3 nm) with an Acton SpectraPro SP2358 spectrometer (grating blaze 1200 nm, 150 lines per mm) and detected with a photodiode (InGaAs, Thorlabs FGA10), then amplified with a transimpedance amplifier (Femto DLPCA-200) and sent to a lock-in amplifier (Stanford Research Systems SR830) that was phase-locked to the waveform generator signal. All spectra were corrected for the absorption of optics in the detection path to obtain the differential photoluminescence signal ΔPL.

**ORCID**

Jana Zaumseil: 0000-0002-2048-217X

Mario Caironi: 0000-0002-0442-4439

ACKNOWLEDGMENTS

This project has received funding from the European Research Council (ERC) under the European Union's Horizon 2020 research and innovation programme (Grant agreement No. 817494 "TRIFECTs" and Grant agreement No. 638059 "HEROIC"). N.F.Z. would like to thank G. Meinusch for valuable technical support and advice throughout the implementation of the CMPL setup.

# Supporting Information

## Table of Contents

## Abundance of SWCNT Species in the Mixed HiPco/PFO Dispersion

**Table S1.** Position of absorption maxima of SWCNT chiralities in the HiPco SWCNT/PFO dispersion, integrated chirality-specific molar absorptivities as determined by Streit *et al.*,[1] and abundance determined from $E_{11}$ peak fits of the absorbance spectrum (see **Figure 1b**).

| Chirality (n,m) | Absorption maximum (nm) | $\int \varepsilon_{11}$ (n,m) d$\lambda$ ($10^5$ M$^{-1}$ cm$^{-1}$) | Abundance (%) |
|---|---|---|---|
| (7,5) | 1043 | 1.67 | 19.7 ± 0.7 |
| (7,6) | 1134 | 2.48 | 21.9 ± 0.4 |
| (8,6) | 1195 | 1.86 | 28.1 ± 0.5 |
| (8,7) | 1283 | 2.59 | 20.0 ± 0.4 |
| (9,7) | 1348 | 1.85 | 10.5 ± 0.5 |

## Raman Spectra

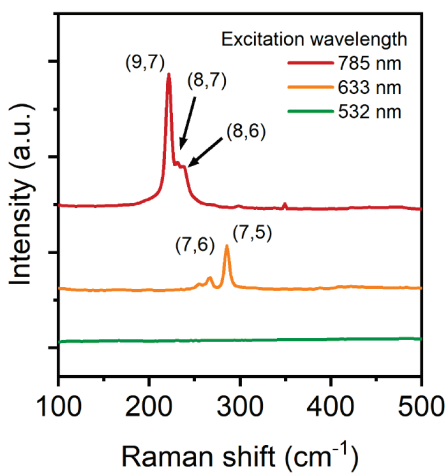

**Figure S1.** Raman spectra of a drop-cast HiPco SWCNT/PFO film in the RBM region. Absence of any RBM peaks for excitation at 532 nm confirms a purely semiconducting network (*i.e.*, no metallic SWCNTs).

**Photoluminescence Excitation-Emission Map**

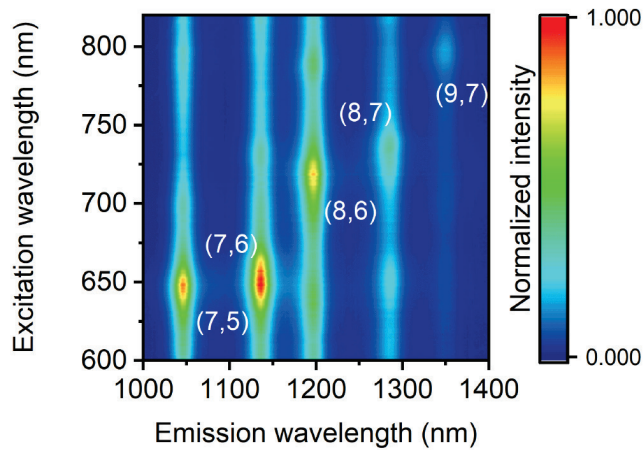

**Figure S2.** Photoluminescence excitation-emission map of a HiPco SWCNT/PFO dispersion.

**Atomic Force Micrographs of SWCNT Networks**

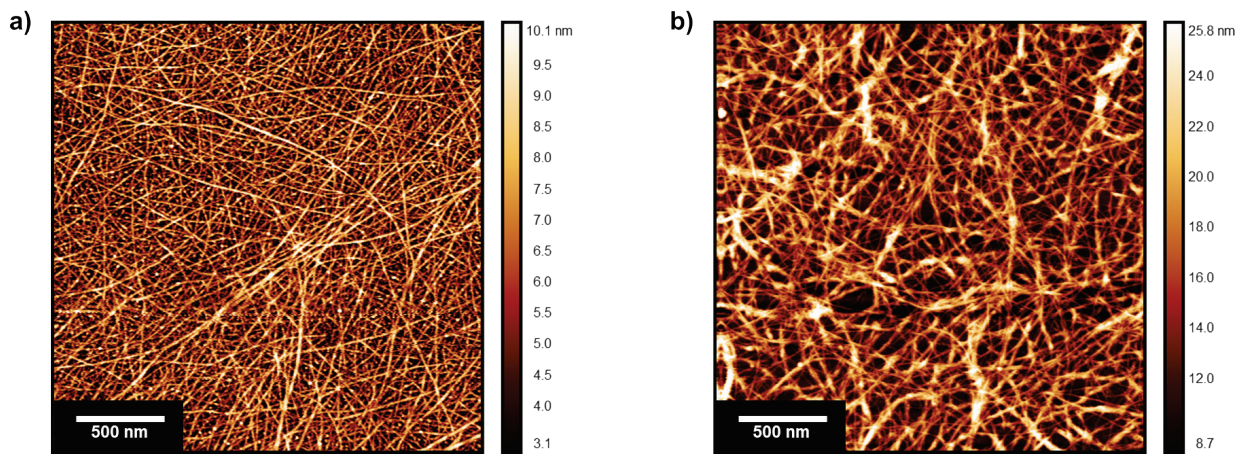

**Figure S3.** AFM images (2.5 x 2.5 µm) of **(a)** a spin-coated (6,5) SWCNT network and **(b)** an aerosol-jet-printed, mixed HiPco/PFO SWCNT network. Lateral scale bars are 500 nm.

## Output Characteristics of (6,5) and Mixed SWCNT Network FETs

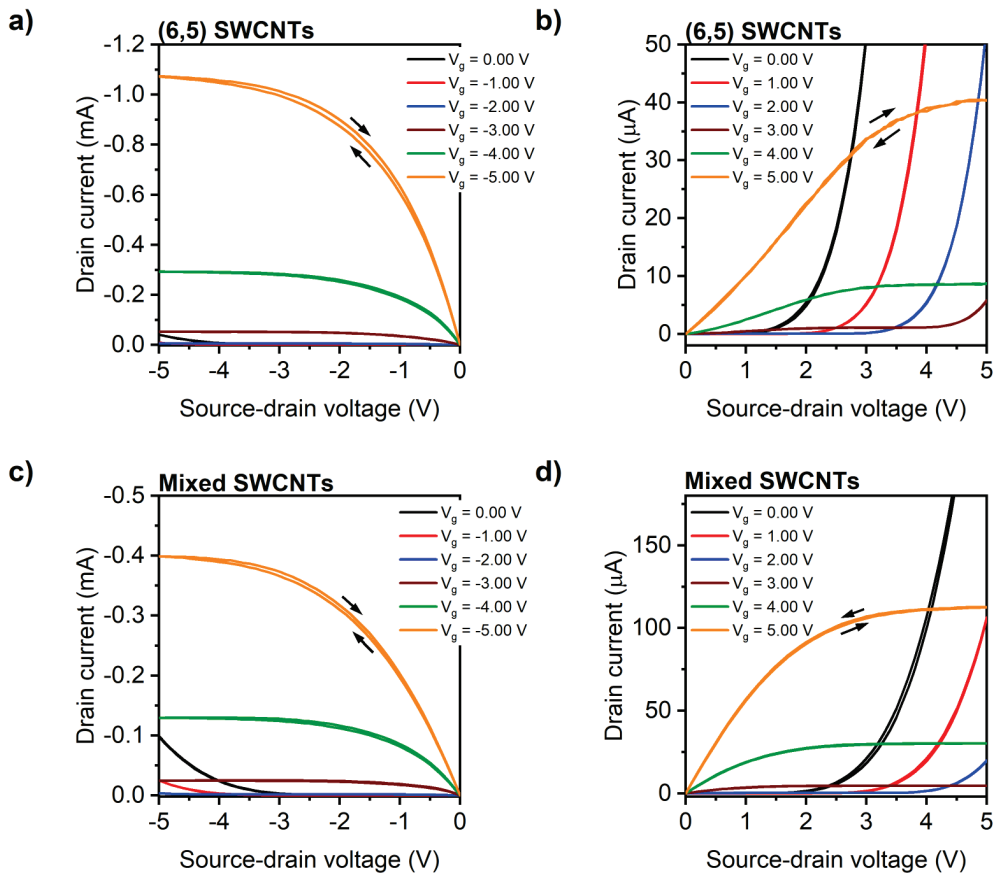

**Figure S4.** Representative output curves of FETs based on **(a,b)** spin-coated (6,5) SWCNTs, and **(c,d)** an aerosol-jet-printed mixed SWCNT network.

**Charge Modulation Spectroscopy Setup**

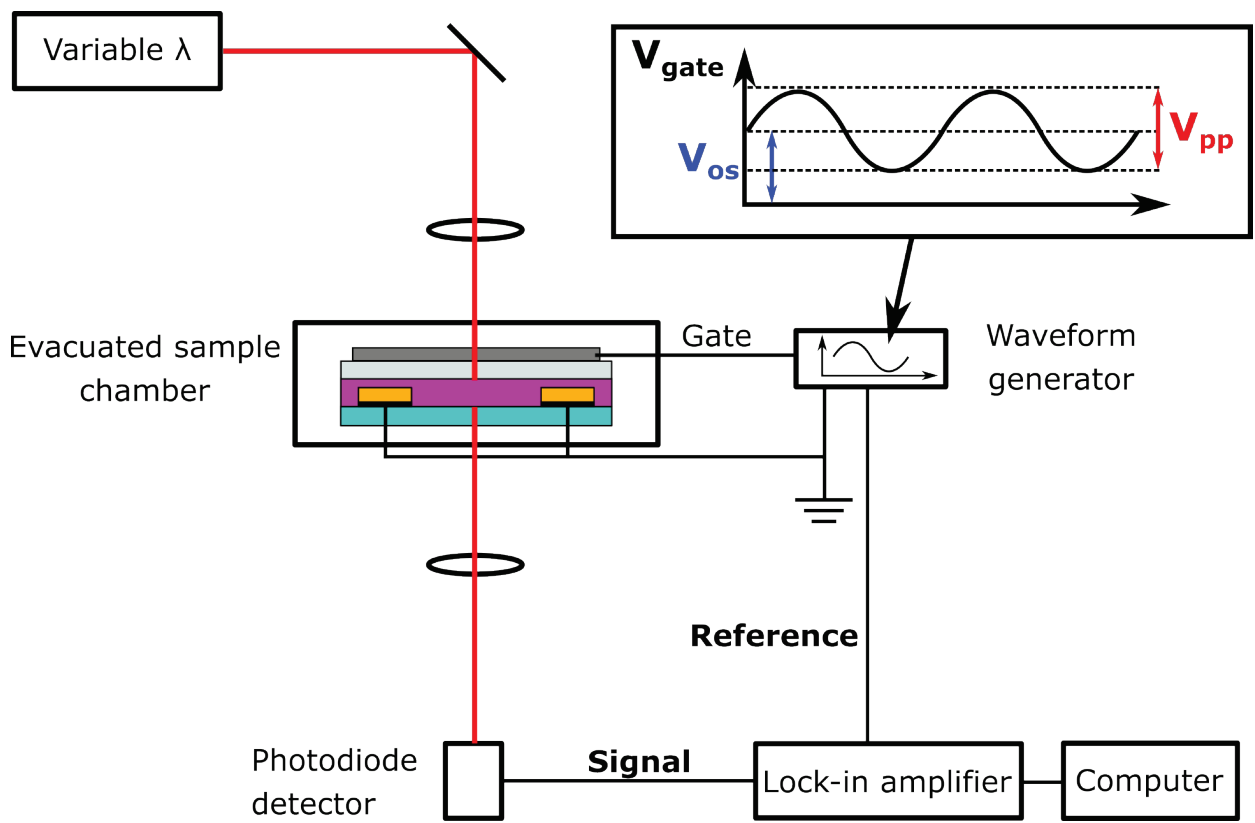

**Figure S5.** Schematic charge modulation absorption spectroscopy (CMS) setup. Note that the reference transmission spectrum T was acquired with no applied bias and a chopper as reference for the lock-in amplifier.

**Complete Set of CMS Data for a (6,5) SWCNT Network FET**

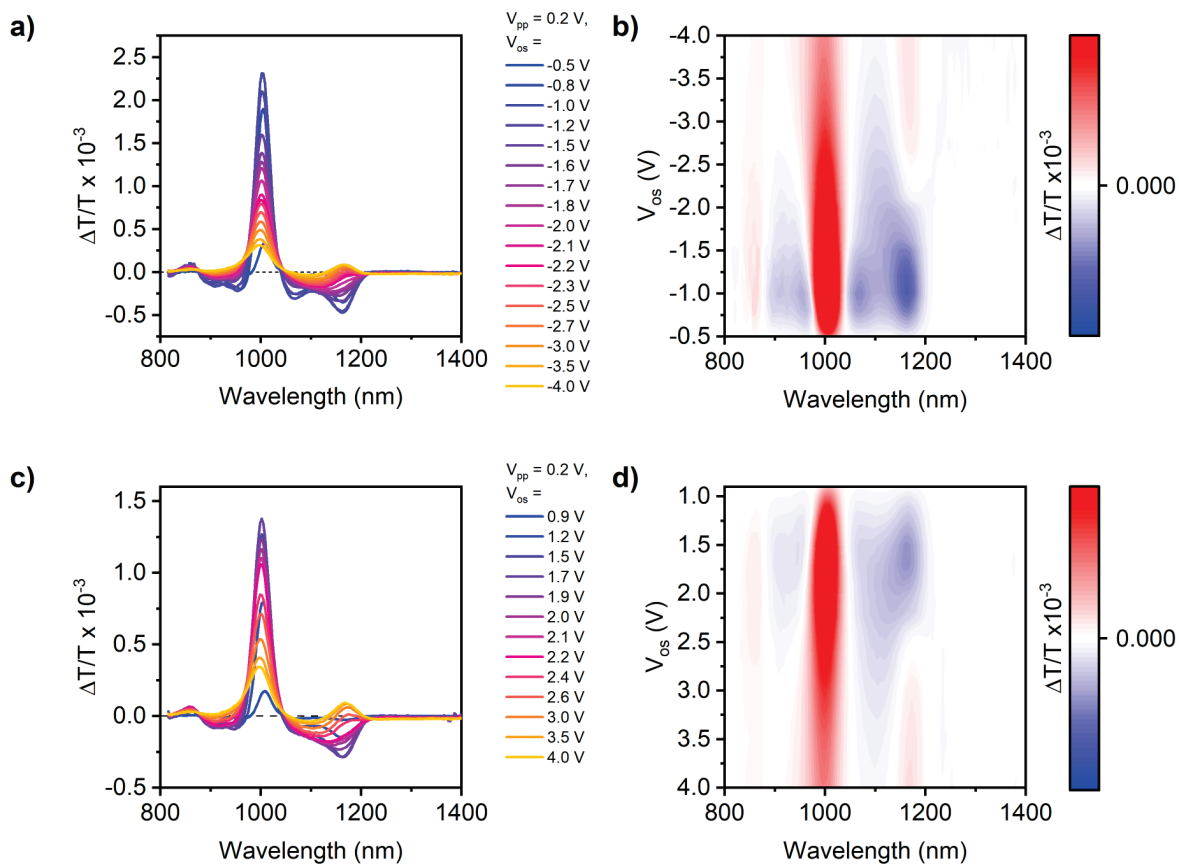

**Figure S6.** Complete dataset of $V_{os}$-dependent CMS spectra ($V_{pp}$ = 0.2 V, modulation frequency 363 Hz) of (6,5) SWCNT network FETs and corresponding colour plots **(a,b)** in hole accumulation and **(c,d)** in electron accumulation.

**Capacitance-Voltage Sweeps of (6,5) and Mixed SWCNT Network FETs**

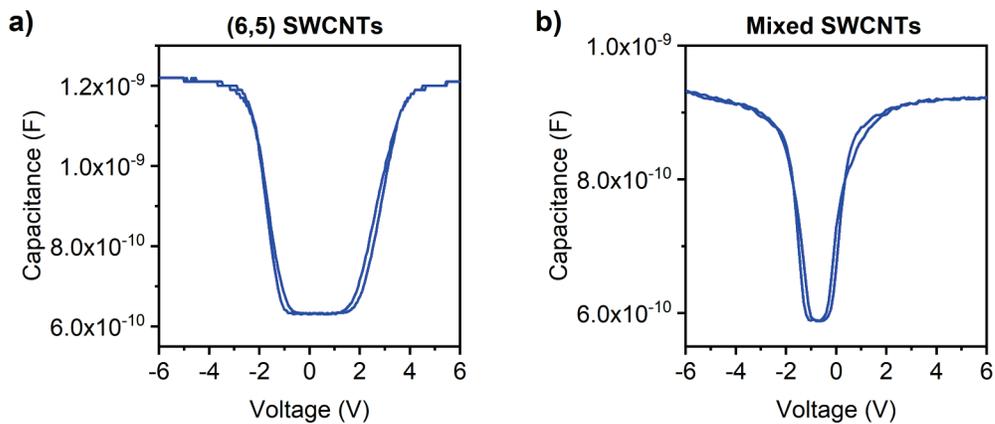

**Figure S7.** Representative capacitance-voltage double sweeps of **(a)** (6,5) SWCNT FETs and **(b)** mixed SWCNT FETs recorded with the usual CMS settings ($V_{pp}$ = 0.2 V, f = 363 Hz, source and drain electrodes shorted).

**Comparison between Static Absorbance Bleaching and CMS Spectra of (6,5) SWCNT Networks**

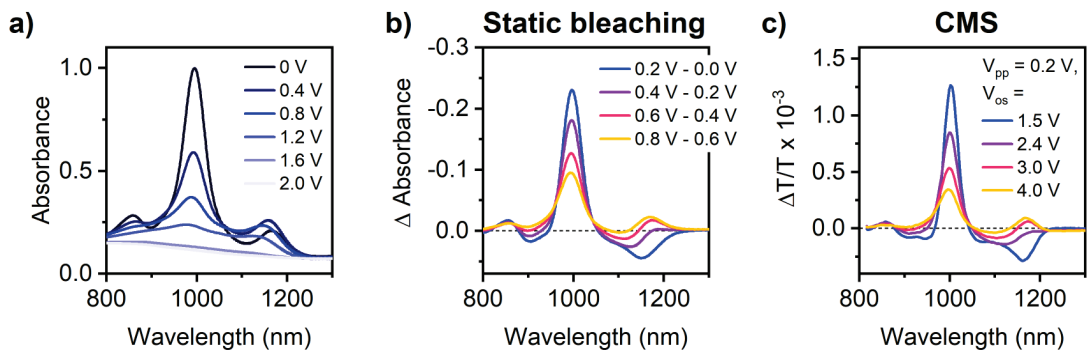

**Figure S8. (a)** Static absorbance bleaching in an electrochromic device based on a thick film of (6,5) SWCNTs. Data reproduced from Berger *et al.*[2] **(b)** Difference spectra (ΔV = 0.2 V) from **(a)** show the decrease in exciton bleaching with increasing charge density as well as the change from negative ("charge-induced absorption") to positive ("bleaching") sign at the wavelength of trion absorption (~1165 nm). **(c)** CMS spectra ($V_{pp}$ = 0.2 V, modulation frequency 363 Hz) of (6,5) SWCNT network FETs for different offset voltages show identical trends compared to the spectroelectrochemical data in **(b)**.

**Additional CMS Cross-Checks, First and Second Harmonic CMS Spectra of (6,5) SWCNTs**

To ensure the correct assignment of the observed transitions and to prevent misinterpretation, we performed additional cross-checks for the CMS spectra. Measurements with identical settings but under different angles of the incident light beam (sample chamber tilted from the usual configuration in which the incident light is orthogonal to the sample/substrate) were performed but did not result in any changes. Hence, any possible interference effects can be excluded.

In previous studies using the CMS method, spectral contributions of electroabsorption (EA) due to the Stark effect were identified.[3,4] We investigated the possible presence of an EA response by detecting the signal at twice the modulation frequency (second harmonic detection scheme, 2ω).[3]

Since the charge density modulation is linearly dependent on the electric field, charge-induced signals should not contribute to the spectra when locking onto the 2ω. The absence of signals in the 2ω spectra (see **Figure S9**) suggests that there is no significant contribution of EA and that the observed CMS signals are purely charge-induced.

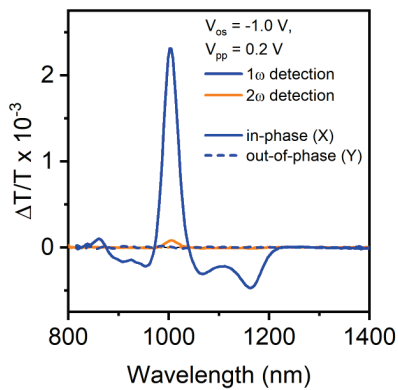

**Figure S9.** CMS spectra of (6,5) SWCNTs ($V_{os}$ = -1.0 V, $V_{pp}$ = 0.2 V) in the first (1ω) and second (2ω) harmonic detection. (X) and (Y) denote the in-phase (solid lines) and out-of-phase (quadrature, dashed lines) component of the signal, respectively.

**Combined vis-nIR CMS Spectra of (6,5) SWCNTs**

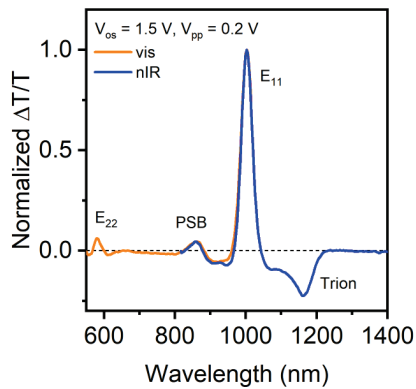

**Figure S10.** Combined CMS spectra of (6,5) SWCNTs in the visible and nIR spectral regions at moderate electron doping level ($V_{os}$ = 1.5 V, $V_{pp}$ = 0.2 V). Spectra are normalized to the $E_{11}$ bleaching signal.

## Normalized Frequency-Dependent CMS Spectra of (6,5) SWCNTs

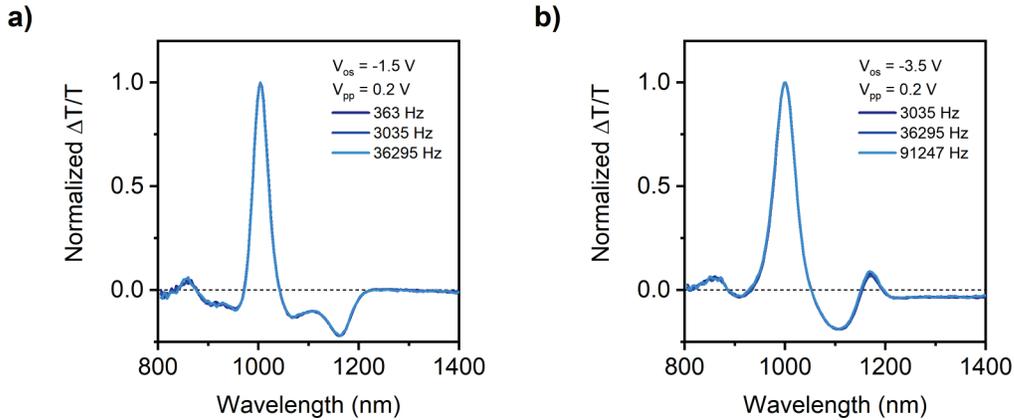

**Figure S11.** Normalized frequency-dependent CMS spectra of (6,5) SWCNTs in the hole accumulation regime **(a)** at moderate doping level ($V_{os}$ = -1.5 V, $V_{pp}$ = 0.2 V) and **(b)** in strong accumulation ($V_{os}$ = -3.5 V, $V_{pp}$ = 0.2 V).

## Gate Voltage-Dependent Linear Mobility of (6,5) and Mixed SWCNT Network FETs

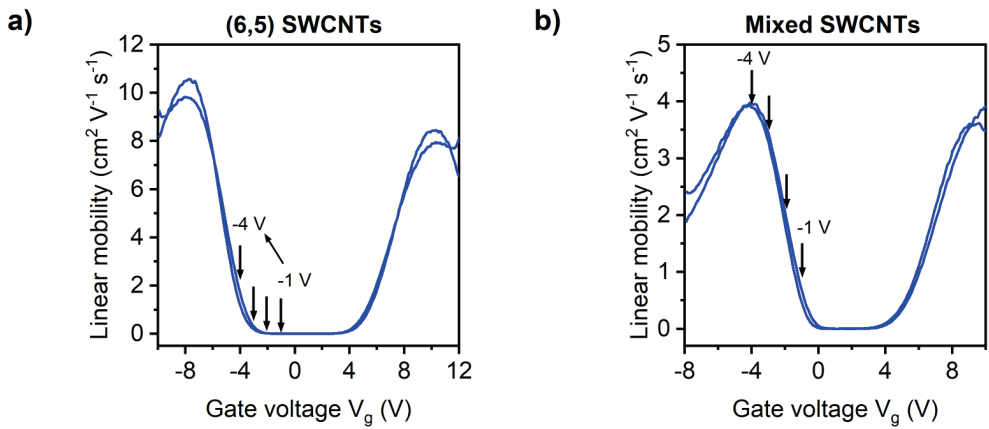

**Figure S12.** Gate voltage-dependent linear mobility ($V_d$ = -0.1 V) of **(a)** a (6,5) SWCNT FET and **(b)** a mixed SWCNT FET. Arrows indicate the voltages that were probed in the frequency-dependent admittance measurements (see **Figure 3c** and **Figure S19c**).

**Frequency-Dependent Capacitance of a (6,5) SWCNT Network FET**

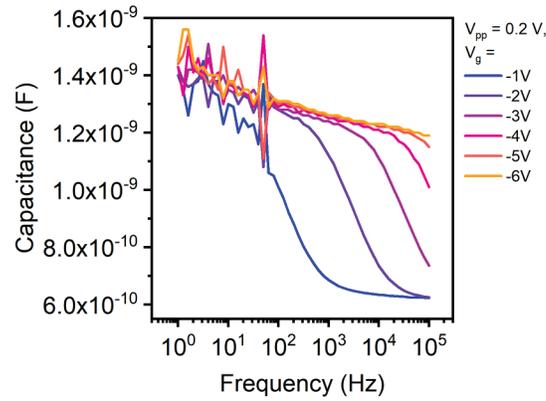

**Figure S13.** Frequency-dependent capacitance of a (6,5) SWCNT network FET for different gate voltages $V_g$ recorded with the usual CMS settings ($V_{pp} = 0.2$ V, source and drain electrodes shorted) show a cut-off with transit time. The slight capacitance decrease observed for all gate voltages is a result of the intrinsic frequency dependence of the PMMA/HfO$_x$ dielectric.

**Phase Plot of the Frequency-Dependent Admittance of a (6,5) SWCNT Network FET**

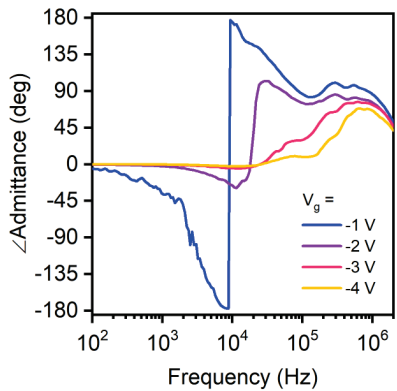

**Figure S14.** Phase plot of the frequency-dependent device admittance (corresponding magnitude of device admittance shown in **Figure 3c**) of a (6,5) SWCNT FET for different gate voltages $V_g$.

**Gate Voltage-Dependent PL of (6,5) SWCNTs**

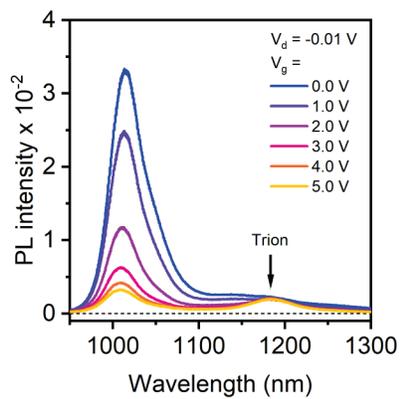

**Figure S15.** Static gate voltage-dependent PL spectra of (6,5) SWCNTs in electron accumulation. The arrow indicates the wavelength of trion emission.

**Charge Modulation PL Spectroscopy Setup**

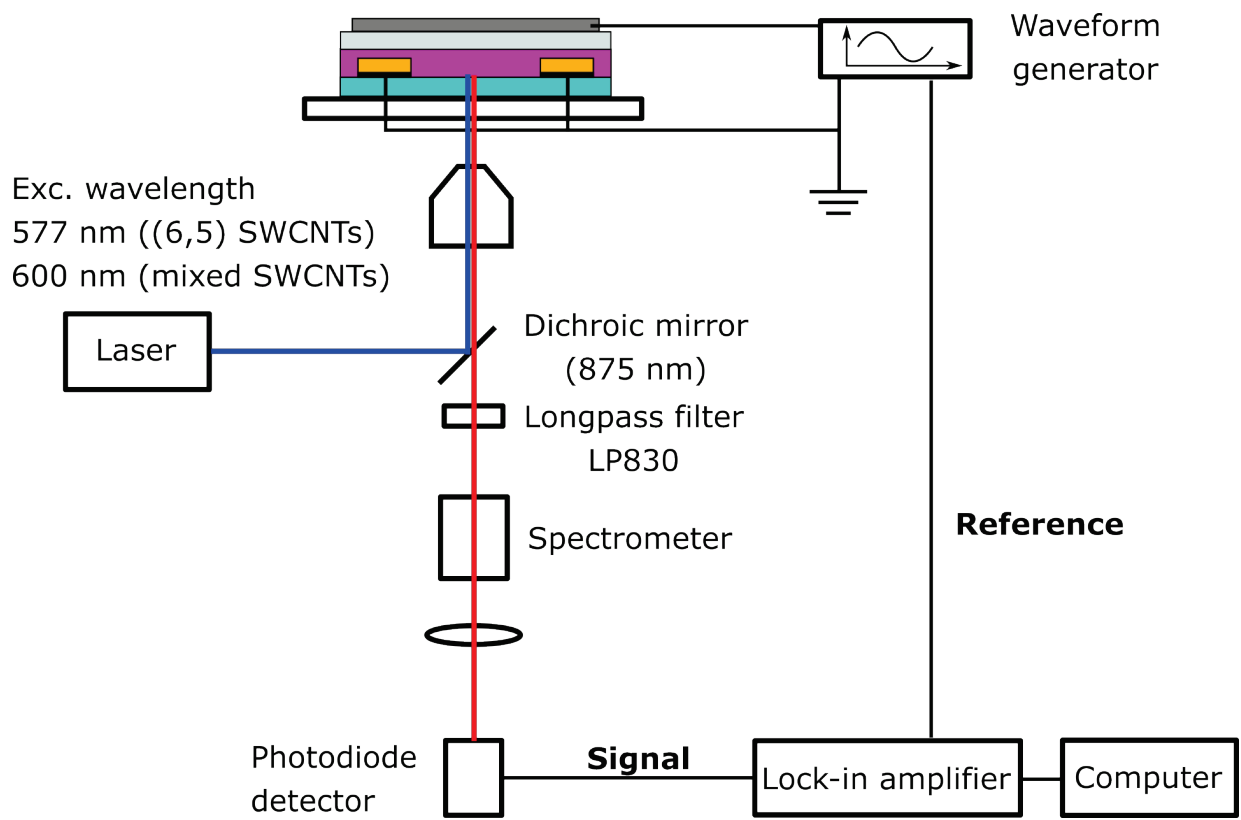

**Figure S16.** Experimental setup for charge modulation photoluminescence (CMPL) spectroscopy. Note that the reference PL spectrum was acquired with no applied bias and by modulating the excitation beam with a mechanical chopper that served as reference for the lock-in amplifier.

## Frequency-Dependent CMPL Spectra of (6,5) SWCNTs

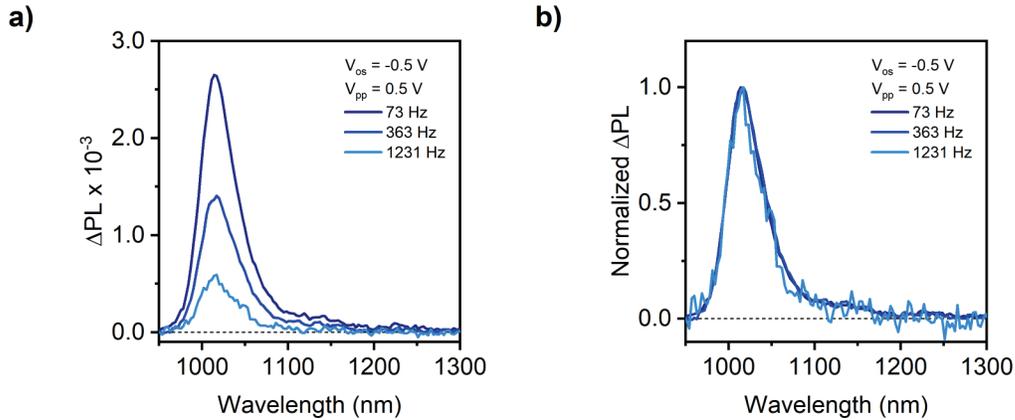

**Figure S17. (a)** Frequency-dependent CMPL spectra of (6,5) SWCNTs in the hole accumulation regime ($V_{os}$ = -0.5 V, $V_{pp}$ = 0.5 V) and **(b)** spectra normalized to the $E_{11}$ quenching signal.

## First and Second Harmonic CMS Spectra of Mixed SWCNTs

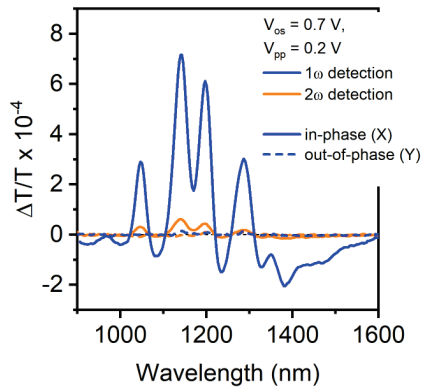

**Figure S18.** CMS spectra of mixed SWCNTs ($V_{os}$ = 0.7 V, $V_{pp}$ = 0.2 V) in the first (1ω) and second (2ω) harmonic detection. (X) and (Y) denote the in-phase (solid lines) and out-of-phase (quadrature, dashed lines) component of the signal, respectively.

**Frequency-Dependent CMS Spectra of Mixed SWCNTs**

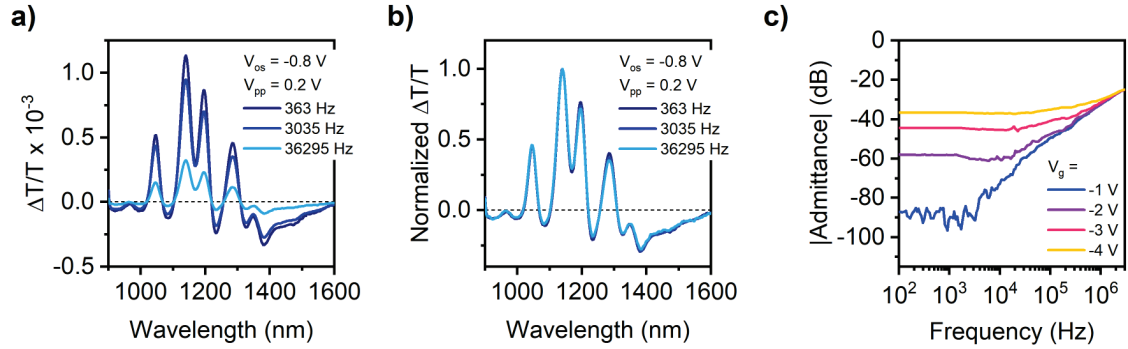

**Figure S19. (a)** Frequency-dependent CMS spectra of mixed SWCNTs in the hole accumulation regime ($V_{os}$ = -0.8 V, $V_{pp}$ = 0.2 V). **(b)** Spectra normalized to the (7,6) SWCNT $E_{11}$ bleaching signal. **(c)** Frequency-dependent magnitude of the device admittance of a mixed SWCNT network FET for different gate voltages $V_g$. The contribution scaling with modulation frequency is due to the source-to-drain capacitance.

**Frequency-Dependent CMPL Spectra of Mixed SWCNTs**

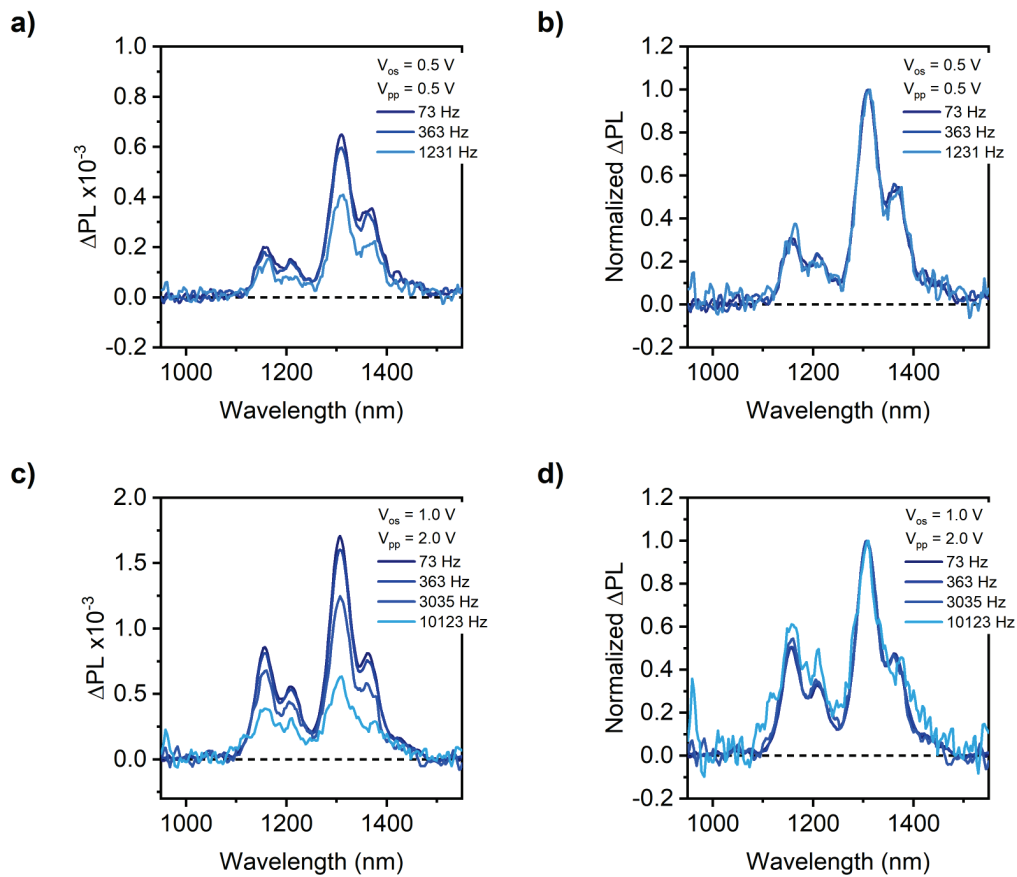

**Figure S20.** Frequency-dependent CMPL spectra of mixed SWCNTs in electron accumulation. **(a)** Spectra acquired at low modulated charge densities ($V_{os}$ = 0.5 V, $V_{pp}$ = 0.5 V). **(b)** Spectra normalized to the (8,7) SWCNT $E_{11}$ quenching signal. **(c)** Spectra acquired at high modulated charge densities ($V_{os}$ = 1.0 V, $V_{pp}$ = 2.0 V). **(d)** Spectra normalized to the (8,7) SWCNT $E_{11}$ quenching signal. Note that the mixed SWCNT network has a different composition (*cf.* **Table S2** and **Figures S21, S22**).

**Absorbance Spectra and Abundance of SWCNT Species in a Mixed Dispersion with Different Composition**

**Table S2.** Absorption maxima of SWCNT chiralities in a different batch of the mixed HiPco/PFO dispersion, integrated chirality-specific molar absorptivities as determined by Streit *et al.*,[1] and abundance determined from $E_{11}$ peak fits of the absorbance spectrum (see **Figure S21a**).

| Chirality (n,m) | Absorption maximum (nm) | $\int \varepsilon_{11}$ (n,m) d$\lambda$ ($10^5$ M$^{-1}$ cm$^{-1}$) | Abundance (%) |
|---|---|---|---|
| (7,5) | 1046 | 1.67 | 10.7 ± 0.7 |
| (7,6) | 1134 | 2.48 | 28.3 ± 0.5 |
| (8,6) | 1196 | 1.86 | 12.5 ± 0.5 |
| (8,7) | 1284 | 2.59 | 35.1 ± 0.5 |
| (9,7) | 1348 | 1.85 | 13.5 ± 0.5 |

**Electrical Characterization of Mixed SWCNT Network FETs and CMPL Spectra with Different Composition**

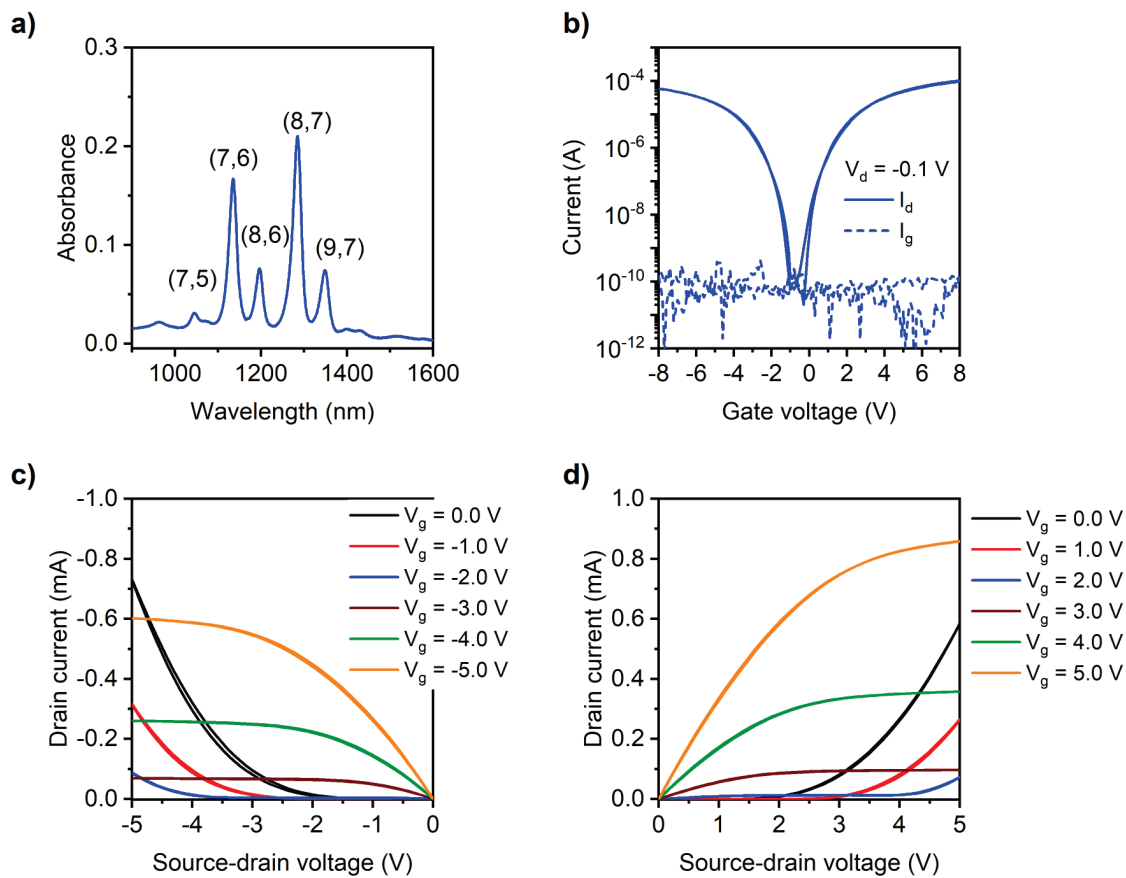

**Figure S21. (a)** Absorbance spectrum of a mixed HiPco/PFO dispersion with different composition. **(b-d)** Representative ambipolar transfer characteristics in the linear regime ($V_d$ = -0.1 V) and output curves of FETs based on random networks of mixed SWCNTs that were aerosol-jet-printed from the dispersion shown in **(a)**.

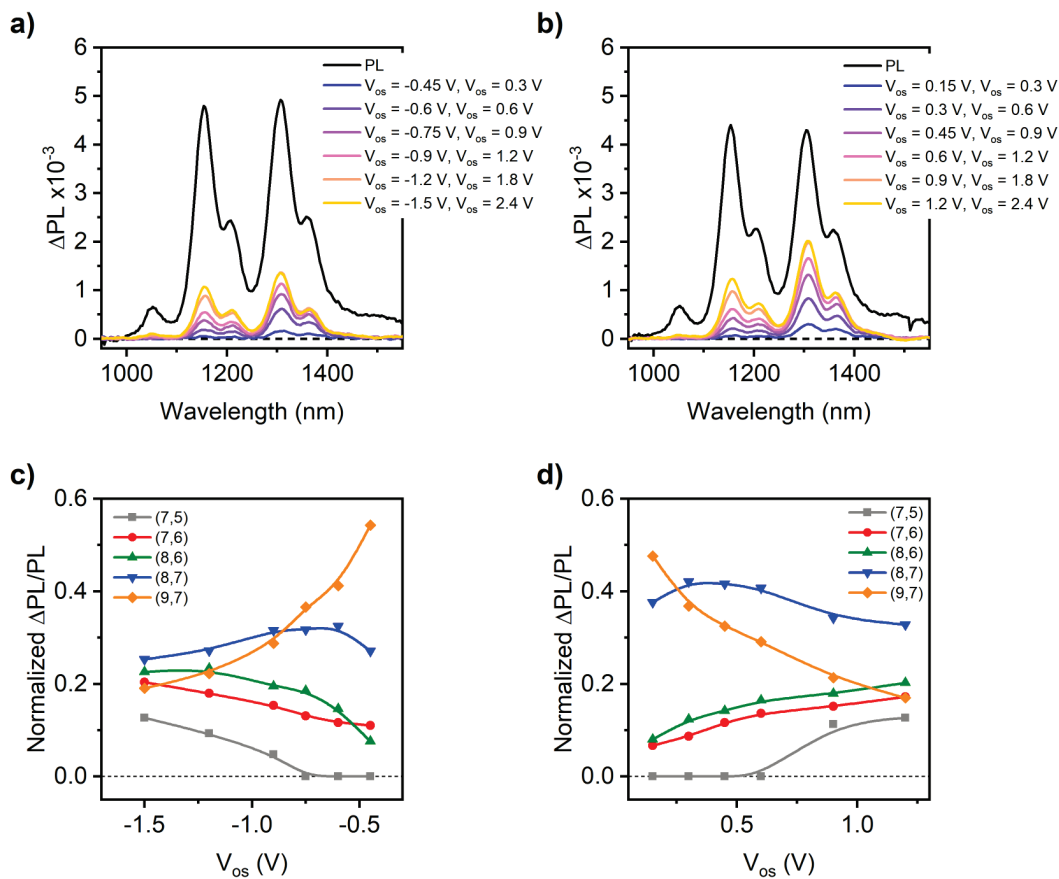

**Figure S22. (a,b)** CMPL spectra of mixed SWCNT FETs with a different network composition in hole and in electron accumulation, respectively, with modulation between off- and on-state of the device. Note that the corresponding transfer curve was slightly shifted to negative voltages (onset for holes, -0.4 V; onset for electrons, 0 V). **(c,d)** Normalized ΔPL/PL for each chirality in the mixed SWCNT networks. Values were determined from peak fits and represent the share of mobile carrier density of this chirality and thus, the current share at varying total charge densities (offset voltages $V_{os}$). Lines are guides to the eye.